\documentclass[twocolumn,superscriptaddress,aps,pra,showpacs,floatfix]{revtex4}

\usepackage{amsmath}
\usepackage{amssymb}
\usepackage{bm}

\usepackage{graphicx}
\usepackage{epsfig}
\usepackage{latexsym}

\usepackage{dcolumn}
\usepackage{graphicx}
\usepackage{epsf}
\usepackage{amsmath}
\usepackage{bm}
\usepackage{maybemath}

\newcommand{\ii}{\mathrm i}

\newcommand{\dd}{\mathrm d}

\begin{document}

\title{Reexamining Black--Body Shifts for Hydrogenlike Ions}

\author{Ulrich D. Jentschura}

\affiliation{Max--Planck--Institut f\"ur Kernphysik,
Postfach 10 39 80, 69029 Heidelberg, Germany}
\affiliation{Institut f\"ur Theoretische Physik,
Philosophenweg 16, 69120 Heidelberg, Germany}

\author{Martin Haas}

\affiliation{Department of Diagnostic Radiology, Medical Physics,
University Hospital Freiburg, 79095 Freiburg, Germany}

\begin{abstract}
We investigate black-body induced energy shifts for low-lying levels of atomic
systems, with a special emphasis on transitions used in current and planned
high-precision experiments on atomic hydrogen and ionized helium.
Fine-structure and Lamb-shift induced black-body shifts are found to increase
with the square of the nuclear charge number, whereas black-body shifts due to
virtual transitions decrease with increasing nuclear charge as the fourth power
of the nuclear charge.
We also investigate the decay width acquired by the
ground state of atomic hydrogen, due to interaction with black-body photons.
The corresponding width is due to an instability against excitation to higher
excited atomic levels, and due to black-body induced ionization. These effects
limit the lifetime of even the most fundamental, a priori absolutely stable,
``asymptotic'' state of atomic theory, namely the ground state of atomic
hydrogen.
\end{abstract}

\pacs{12.20.Ds, 31.15.-p, 42.50.Hz}

\maketitle

%
%
\section{Introduction}

Energy shifts of atomic levels due to interactions with black-body radiation
are primarily important for Rydberg states, where large dipole polarizabilities
enhance the magnitude of the effect~\cite{GaCo1979,FaWi1981,Ga1994}, but at the
current rate of advance of high-precision spectroscopy, its importance for the
correct realization of frequency standards has also been stressed in the
literature~\cite{ItLeWi1981,ItLeWi1982,HoHa1984,VaAu1989}.  In particular, the
work of Farley and Wing~\cite{FaWi1981} was mainly focused on Rydberg states,
which were accessible to high-precision spectroscopy at the time. In contrast,
we here focus on the ground state and on the $2S$ state of hydrogen and
hydrogenlike ions. The transition frequencies in these systems have only
recently come within reach of high-precision laser spectroscopy, due to 
the development of frequency combs. In all cases studied by Farley and
Wing~\cite{FaWi1981}, the Lamb shift and fine structure shifts could be
neglected, but this does not hold for the transitions studied here.

Ionization by black-body radiation also
is an important effect for excited atomic states.
The ionization process leads to a finite width of the 
states (resonances), with the energy acquiring a (small)
imaginary part. The sign of the imaginary part is negative,
and we can write the black-body induced energy shift 
$\Delta {\cal E}_{\rm bb}$ as
\begin{equation}
\Delta {\cal E}_{\rm bb} =
\Delta E_{\rm bb} - 
\ii \frac{\Delta \Gamma_{\rm bb}}{2}
\end{equation}
where $\Delta E_{\rm bb}$ is the real part of the energy shift,
and $\Delta \Gamma_{\rm bb}/h$ is the black-body induced width.

Several peculiarities characterize the black-body induced 
radiative shift. First of all, we recall that two virtual 
processes contribute: one where the atom absorbs a black-body photon
and then returns to the reference state by emission into the 
same mode of the electromagnetic field from which 
a photon had been absorbed, and another one where the sequence 
of absorption and emission processes is reversed.
This is very much analogous to the so-called ac Stark shift
(see, e.g.,~\cite{Sa1994Mod,HaJeKe2006}) that an atom 
feels in a laser field, but with the difference that 
the black-body radiation is isotropic, and that the polarization
vectors are equally distributed among all possible directions in space.
For the ground state being the reference state, the ``width''
(imaginary part of the energy shift)
is exclusively generated by the absorption-first channel,
where one of the propagator denominators becomes resonant
and the virtual state takes the role of the final state
of the ionization or excitation process~\cite{HaEtAl2006}.
Note that recently~\cite{EsSo2008}, a finite width has been
predicted for the ground state of hydrogen at very high 
temperatures on the basis of field-theoretical considerations
[see Eq.~(23) of Ref.~\cite{EsSo2008}].

One observation, potentially of fundamental interest, 
substantiated here by a concrete calculation,
is that even the $1S$ state of atomic hydrogen, 
acquires a finite width when the hydrogen atom 
is exposed to black-body radiation, although
the ground state of the most fundamental atomic system
is otherwise assumed to be the perfect ``asymptotic state'' 
used in all $S$-matrix type calculations regarding the 
quantum electrodynamic (QED) shifts of energy 
levels~\cite{GMLo1951,Su1957,MoPlSo1998}.
However, before we come to an analysis of this 
effect, we briefly revisit the evaluation of the 
real part of the black-body induced energy shifts, 
for the $1S$--$2S$ transition in 
hydrogen~\cite{NiEtAl2000,FiEtAl2004} and ionized helium
(see Sec.~\ref{realparts}).
We then analyze the imaginary parts of the black-body
induced energy shifts in Sec.~\ref{imagparts}.
Conclusions are drawn in Sec.~\ref{conclu}.

%
%
\section{Real Parts of Black-body Radiation Induced Energy Shifts}
\label{realparts}

For clarity, we keep all factors of $h$, $c$ and $\epsilon_0$
in all calculations. The Boltzmann constant is denoted $k_B$.
We recall that the energy distribution per frequency interval
$\dd \nu$ of black-body radiation is
\begin{equation}
\label{bbspec}
\rho(\nu) \, \dd \nu =
\frac{8 \, \pi \, h \, \nu^3}{c^3} \,
\left[ \exp\left( \frac{h \nu}{k_B T} \right) - 1 \right]^{-1} 
\dd \nu \,,
\end{equation}
which is connected to the time-averaged square of the electric field 
$\vec e^{\,2}(\nu)$ at frequency~$\nu$ (in units of the square of the 
electric field strength per frequency) as 
\begin{equation}
\label{spectre}
\vec e^{\,2}(\nu) \, \dd \nu = 
\frac{1}{\epsilon_0} \, \rho(\nu) \, \dd \nu\,.
\end{equation}
We write the Cartesian coordinates as $x^i$ and rewrite the 
dipole polarizability of an $nS$ state to take the 
isotropic character of the polarization vectors of black-body radiation 
into account [cf.~Eq.~(20) of Ref.~\cite{HaJeKe2006}].
Then, the black-body energy shift 
$\Delta E_{\mathrm{bb}}( nS )$ is given by the formulas
\begin{subequations}
\begin{align}
\label{pol}
P_{\nu}(nS) =& 
\sum\limits_\pm 
\sum_{i=1}^3 
\frac13 
\left< \! nS \left| x^i \frac{1}{H - E(nS) \pm h \nu} x^i
\right| nS \! \right>, \\[2ex]
\label{ebb}
\Delta E_{\mathrm{bb}}( nS ) =& - \frac{ e^2 }{2} \,
({\mathrm{P.V.}}) \int\limits_0^\infty \dd \nu\,
\vec e^{\,2}(\nu) \, P_{\nu}(nS) \,,
\end{align}
\end{subequations}
where $E(nS)$ is the energy of the reference $nS$ state.
The latter integral actually has to be taken 
as a principal-value (P.V.) integral because the propagator denominators
in (\ref{pol}) can become singular when a black-body photon hits an atomic 
resonance (see the discussion below in Sec.~\ref{imagparts}). 

For the $1S$ state, the properties of the dipole polarizability 
$P_{\nu}(1S)$ are well understood~\cite{GaCo1970}, and 
$P_{\nu}(1S)$ can be evaluated using an entirely nonrelativistic 
approximation~\cite{Pa1993}, and $H$ in \eqref{pol} 
may be replaced by the Schr\"{o}dinger Hamiltonian.
To see why that is the case, let us consider the 
typical photon frequencies at which the black-body photon energy
peaks. Ignoring the ``$-1$'' in the denominator of \eqref{spectre},
it is easy to derive the result
\begin{equation}
\label{numax}
\nu_{\rm max} \approx \frac{3 k_B \, T}{h}
\end{equation}
for the frequency at which $\rho(\nu)$ assumes its maximum value.
We consider a temperature range from $4 \, {\rm K} \dots 300 \, {\rm K}$.
In this temperature range, 
the frequency $\nu_{\rm max}$ varies from $2.5 \times 10^{11} \, {\mathrm{Hz}}$ 
to $1.9 \times 10^{13} \, {\mathrm{Hz}}$, which is well below the 
Rydberg constant expressed in frequency units, 
$R_\infty \, c = 3.298 \times 10^{15} \, {\mathrm{Hz}}$,
which defines the frequency range for the virtual excitations to 
$P$ levels that are decisive for the evaluation of the 
dipole polarizability. For the ground state acting as the 
reference state, we may therefore even ignore the 
terms $\pm h \nu$ in the propagator denominators in 
\eqref{pol} and replace the dynamic by the static polarizability,
which can be evaluated easily and is used here in the nonrelativistic
approximation (cf.~Ref.~\cite{Ya2003})
\begin{equation}
\label{static}
P_{0}(1S) = \frac{9}{2} \,
\left( \frac{\hbar}{m_{\mathrm e} \, c} \right)^2 \, 
\frac{1}{(Z \alpha)^4 \, m_{\mathrm e} c^2} \,.
\end{equation}
The dipole polarizability defined in Eq.~(\ref{pol})
parametrically scales as $Z^{-4}$, because the 
energy differences in the propagator denominators
scale with $Z^2$, whereas the two dipole transition matrix elements in 
the numerator each acquire a factor $1/Z$
(the atom becomes ``smaller'' by a factor $Z^{-1}$ as the 
nuclear charge number increases). 
We finally obtain the approximation
\begin{equation}
\label{deltae1Sbb}
\Delta E_{\mathrm{bb}}( 1S ) \approx  
- \frac{ 3 \pi^3 \, k_B^4 }%
{ 5 \, \alpha^3 \, m_{\mathrm e}^3 \, c^6} \,
\frac{T^4}{Z^4}.
\end{equation}
The corresponding data in Table~\ref{table1} are in excellent 
agreement with this approximation.

\begin{table}
\begin{center}
\begin{minipage}{0.9\linewidth}
\begin{center}
\caption{\label{table1} Black-body energy shift of the 
ground state of atomic hydrogen and ionized atomic helium,
at a temperature of $T = 4 \, {\mathrm K}$,
$T = 77 \, {\mathrm K}$ and $T = 300 \, {\mathrm K}$.
The shift $\Delta E_{\mathrm{bb}}( 1S )$ is obtained 
by numerical integration according to Eq.~\eqref{ebb} and divided 
by the Planck constant in order to be expressed in frequency units.}
\begin{tabular}{ccr@{\hspace{0.0cm}}l}
\hline
\hline
\multicolumn{1}{c}{\rule[-2mm]{0mm}{6mm} Nucl. charge} &
\multicolumn{1}{c}{Temperature} &
\multicolumn{2}{c}{$h^{-1} \, \Delta E_{\mathrm{bb}}( 1S )$} \\
\multicolumn{1}{c}{\rule[-2mm]{0mm}{6mm} number} & & & \\
\hline
\rule[-2mm]{0mm}{6mm}
$Z = 1$ &
$4 \, {\rm K}$ &
$-1.$&$22 \times 10^{-9}\, {\mathrm{Hz}}$ \\
\rule[-2mm]{0mm}{6mm}
$Z = 1$ &
$77 \, {\rm K}$ &
$-1.$&$68 \times 10^{-4}\, {\mathrm{Hz}}$ \\
\rule[-2mm]{0mm}{6mm}
$Z = 1$ &
$300 \, {\rm K}$ &
$-3.$&$88 \times 10^{-2} \, {\mathrm{Hz}}$ \\
\rule[-2mm]{0mm}{6mm}
$Z = 2$ &
$4 \, {\rm K}$ &
$-7.$&$65 \times 10^{-12}\, {\mathrm{Hz}}$ \\
\rule[-2mm]{0mm}{6mm}
$Z = 2$ &
$77 \, {\rm K}$ &
$-1.$&$05 \times 10^{-5}\, {\mathrm{Hz}}$ \\
\rule[-2mm]{0mm}{6mm}
$Z = 2$ &
$300 \, {\rm K}$ &
$-2.$&$42 \times 10^{-3}\, {\mathrm{Hz}}$ \\
\hline
\hline
\end{tabular}
\end{center}
\end{minipage}
\end{center}
\end{table}

A subtle point concerns the dipole 
polarizability of excited states~\cite{Ya2003}: While for the $1S$ state,
it is entirely sufficient to approximate $H$ by the Schr\"{o}dinger 
Hamiltonian, this is not the case for the $2S$ state and for 
higher excited $nS$ states with $n > 2$. Indeed, the 
$2P_{1/2}$ state is displaced from $2S$ only by the Lamb shift, 
and the $2P_{3/2}$ state is displaced 
in energy only by the fine-structure, and yet, 
the dipole transition matrix elements between the $2S$ and the 
$2P$ states are manifestly nonvanishing. The dipole transition
matrix elements which enter the numerators for the dipole polarizability,
can still be taken in a nonrelativistic approximation,
but the denominators must be adjusted for the Lamb shift and the 
fine-structure effects. In summary, we have for an $nS$ level with 
$n \geq 2$,
\begin{align}
\label{PnS}
& P_{\nu}(nS) 
\nonumber\\
& = \frac13 \, \sum\limits_\pm \sum_{i=1}^3 \sum_{\mu = -1/2}^{1/2} 
\frac{ \left| \left< nS_{1/2} (m \! = \! \frac12) \left| x^i \right| 
nP_{1/2} (m \! = \! \mu) \right> \right|^2 }{ E(nP_{1/2}) - E(nS) \pm h \nu}
\nonumber\\[2ex]
& + \frac13 \, \sum\limits_\pm \sum_{i=1}^3 \sum_{\mu = -3/2}^{3/2} 
\frac{ \left| \left< nS_{1/2} (m \! = \! \frac12) \left| x^i \right| 
nP_{3/2} (m \! = \! \mu) \right> \right|^2 }{ E(nP_{3/2}) - E(nS) \pm h \nu}
\nonumber\\[2ex]
& + \frac13 \, \sum\limits_\pm \sum_{i=1}^3 \sum_{\mu = -1}^{1} 
\sum_{n'\neq n}
\frac{ \left| \left< nS (m \! = \! 0) \left| x^i \right| 
n' P (m \! = \! \mu) \right> \right|^2 }{ E(n'P) - E(nS) \pm h \nu } \,.
\end{align}

Here, the $| nS_{1/2} \rangle$, the $| nP_{1/2} \rangle$ and the 
$| nP_{3/2} \rangle$ states can be approximated 
by Schr\"{o}dinger--Pauli wave functions,  i.e.~the radial part 
is taken in the nonrelativistic approximation, but the 
angular part is given by a spinor (two-component) function 
$\chi_\mu^\kappa(\hat{r})$ defined as in~\cite{Ro1961};
$\kappa = 2 \, (l - j) \, (j + 1/2)$ is the 
Dirac angular quantum number, with $l$ being the orbital angular momentum
quantum number and $j$ the total electron angular momentum,
and $\mu \in \{ -(j+1/2), \dots, j+1/2 \}$ 
is the magnetic projection (half-integer).
The states $nP$ and $2S$, by contrast, are plain nonrelativistic
Schr\"{o}dinger eigenstates with an angular part of the 
form $Y_{lm}(\hat{r})$, where $l$ is the orbital angular momentum,
and $m \in \{ -l, \dots, l \}$ is the magnetic projection (integer). 

In the following, we concentrate on the case $n = 2$,
i.e. on the black-body shift of the $2S$ state.
The key to finding an analytic approximation to $\Delta E_{\mathrm{bb}}(2S)$
is the following. We observe that for most cases of interest studied
here, the condition
\begin{equation}
\label{hierarchy}
E_{\mathrm{fs}} \ll h \, \nu_{\mathrm{max}} \ll h \, R_\infty \,c \,,
\end{equation}
is fulfilled, where
$R_\infty \, c \approx 3.289 \times 10^{15} \, {\mathrm{Hz}}$
is the Rydberg constant expressed in frequency units.
This hierarchy means that we can make different approximations
for the first two as opposed to the third term on the
right-hand side of \eqref{PnS}. Namely, the relevant 
frequency range of the black-body radiation is large as compared
to the fine-structure and the Lamb shift, but small in 
comparison to the frequencies corresponding to transition  with 
a change in the principal quantum number.
For the last term, of \eqref{PnS}, we can thus make the 
static approximation and obtain
\begin{align}
\label{PnSprime}
P'_0(2S) 
=& \; \lim_{\nu \to 0} 
\sum\limits_\pm \sum_{i=1}^3 \sum_{\mu = -1}^{1} 
\sum_{n' = 3}^\infty
\nonumber\\[2ex]
& \; \quad \times \frac13 \, 
\frac{ \left| \left< 2S (m=0) \left| x^i \right| 
n' P (m = \mu) \right> \right|^2 }{ E(n'P) - E(2S) \pm h \nu } 
\nonumber\\[2ex]
=& \; 120 \,
\left( \frac{\hbar}{m_{\mathrm e} \, c} \right)^2 \, 
\frac{1}{(Z \alpha)^4 \, m_{\mathrm e} c^2} \,,
\end{align}
where we reemphasize that the sum over $n'$ starts from $n' = 3$.
The corresponding energy shift is 
\begin{equation}
\label{Deltae2Sbbprime}
\Delta E'_{\mathrm{bb}}( 2S ) \approx  
- \frac{ 16  \pi^3 \, k_B^4 }%
{ \alpha^3 \, m_{\mathrm e}^3 \, c^6} \,
\frac{T^4}{Z^4}.
\end{equation}
For the black-body shift corresponding to the first two terms 
on the right-hand side of~\eqref{PnS}, we first recall
the fine-structure interval and the Lamb shift in
leading order,
\begin{subequations}
\begin{align}
& E_{\mathrm{fs}} = E(2P_{3/2}) - E(2S_{1/2}) = 
\frac{(Z\alpha)^4 \, m_{\mathrm e} \, c^2}{32} \,,
\\[2ex]
& E_{\mathrm L} = E(2S_{1/2}) - E(2P_{1/2}) = 
\frac{\alpha}{8 \pi} \, (Z\alpha)^4 \, m_{\mathrm e} \, c^2 \,
\\[2ex]
& \; \times \left\{ \frac43 \, \ln\left[ (Z\alpha)^{-2} \right] + 
\frac43 \, \left( \ln k_0 (2P) - \ln k_0(2S) \right) + \frac{91}{90} 
\right\} \,.
\nonumber
\end{align}
\end{subequations}
Here, we make the opposite approximation
and expand the terms for large $\nu$,
\begin{align}
\label{PnSdprime}
& P''_{\nu}(2S) 
\nonumber\\[2ex]
& = \frac13 \, \sum\limits_\pm \sum_{i=1}^3 \sum_{\mu = -1/2}^{1/2} 
\frac{ \left| \left< 2S_{1/2} (m \! = \! \frac12) \left| x^i \right| 
2P_{1/2} (m \! = \! \mu) \right> \right|^2 }{ E(2P_{1/2}) - E(2S) \pm h \nu}
\nonumber\\[2ex]
& + \frac13 \, \sum\limits_\pm \sum_{i=1}^3 \sum_{\mu = -3/2}^{3/2} 
\frac{ \left| \left< 2S_{1/2} (m \! = \! \frac12) \left| x^i \right| 
2P_{3/2} (m \! = \! \mu) \right> \right|^2 }{ E(2P_{3/2}) - E(2S) \pm h \nu}
\nonumber\\[2ex]
& \approx
-\frac{1}{\nu^2} \,
\frac{3 E_{\mathrm{fs}}}{\pi^2 (Z\alpha)^2 m_{\mathrm e}^2 c^2}
+\frac{1}{\nu^2} \,
\frac{3 E_{\mathrm L}}{2 \pi^2 (Z\alpha)^2 m_{\mathrm e}^2 c^2} \,.
\end{align}

\begin{figure*}[htb]
\includegraphics[width=0.55\linewidth]{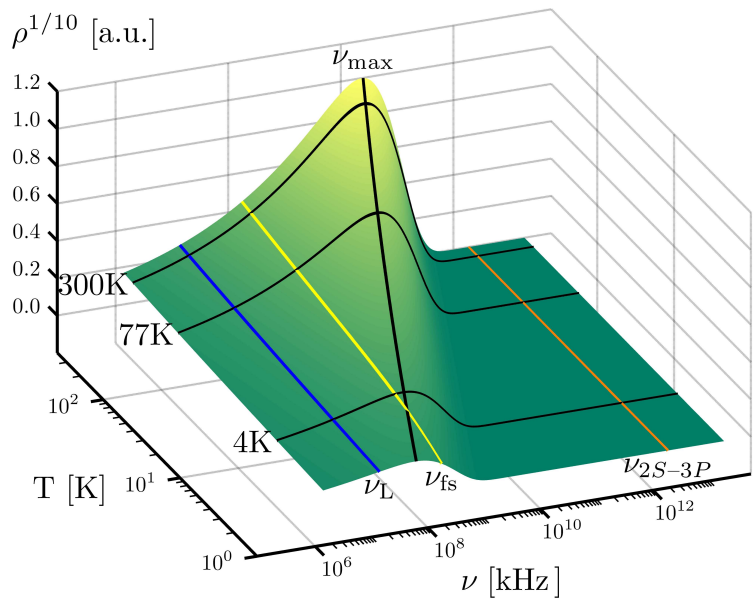}
\caption{\label{fig1}
The black-body spectrum given by \eqref{bbspec} 
is displayed as a function of the frequency $\nu$
and of the temperature $T$. In order to visualize the 
spectrum over a wide range of temperatures, we 
plot $\rho^{1/10}$, and we use arbitrary units (a.u.), normalized
so that $\rho$ assumes a value of unity at its 
maximum (as a function of $\nu$) for $T = 400 \, {\rm K}$.
The maximum of the blackbody spectrum as a function of
the frequency at given temperature is
denoted as $\nu_{\rm{max}}$ [see also Eq.~\eqref{numax}]. 
The relation of $\nu_{\rm{max}}$
to the relevant atomic frequencies,
as immediately discernible from the plot,
illustrates why the approximation \eqref{Deltae2Sbbprime}
and \eqref{Deltae2Sbbdprime} is not applicable to ionized helium
at $T = 4\, {\rm K}$. The frequency $\nu_{2S-3P}$ is the frequency
for excitation of the $2S$ state to the $3P$ state for ionized helium.}
\end{figure*}

Using \eqref{spectre}, we can integrate integrate over the 
black-body spectrum and obtain 
\begin{align}
\label{Deltae2Sbbdprime}
& \Delta E''_{\mathrm{bb}}(2S) = 
\frac{\pi \alpha^3 k_B^2}{8 m_{\mathrm e} \, c^2} \, T^2 \, Z^2 
- \frac{\alpha^4 k_B^2}{4 m_{\mathrm e} \, c^2} \, 
\left\{ \frac43 \, \ln\left[ (Z\alpha)^{-2} \right] 
\right.
\nonumber\\[2ex]
& \; \left. 
+ \frac43 \, \left( \ln k_0 (2P) - \ln k_0(2S) \right) + \frac{91}{90} 
\right\} \, T^2 \, Z^2.
\end{align}
The approximation then is
\begin{equation}
\label{approx2S}
\Delta E_{\mathrm{bb}}(2S) \approx
\Delta E'_{\mathrm{bb}}(2S) +
\Delta E''_{\mathrm{bb}}(2S) \,.
\end{equation}
From Table~\ref{table2} we see that the approximation 
holds for all cases studied except 
for the shift of the $2S$ state
for ionized helium ($Z = 2$) at $4\,{\mathrm K}$.
The inequality \eqref{hierarchy} is not fulfilled in this case,
because $\nu_{\mathrm{max}} = 2.5 \times 10^{11} \, {\mathrm{Hz}}$
at $4\, {\mathrm K}$
whereas $E_{\mathrm{fs}}/h  = 1.8 \times 10^{11} \, {\mathrm{Hz}}$
for $Z = 2$ (see also Fig.~\ref{fig1}). For $Z = 1$, at the same temperature, 
the inequality $E_{\mathrm{fs}} \ll h \, \nu_{\mathrm{max}}$ 
is better fulfilled, and this affords an explanation
for the fact that the approximation works well in the latter case.

One would naively assume that the black-body shifts 
should decrease with the nuclear charge, because of the 
$Z^{-4}$ scaling of the polarizability. 
However, that is not the case. 
The formulas~\eqref{Deltae2Sbbprime} and \eqref{Deltae2Sbbdprime}
indicate that there are two competing effects for the 
shift of the $2S$ state, one which is related to 
virtual transitions with a change in the principal quantum 
number [see Eq.~\eqref{Deltae2Sbbprime}], and another one which is 
related to fine-structure and Lamb shift transitions
[see Eq.~\eqref{Deltae2Sbbdprime}]. The former scales with 
$T^4 \, Z^{-4}$, as expected, but the latter scales as $T^2 \, Z^2$,
and increases with the nuclear charge number (somewhat
counterintuitively). We would thus like to 
refer to this as an anomalous scaling.
This particular behavior, in connection with planned
experiments~\cite{HeEtAl2008}, has been a major motivation for 
carrying out the calculation report here.

Incidentally, we note that the black-body shifts at $T = 300\,{\rm K}$
are by several orders of magnitude 
larger than other non-standard, 
non-resonant effects (``accuracy limits'') for two-photon 
spectroscopy recently discussed in Refs.~\cite{JeMo2002,LaScSoPl2007}.
Note that after leaving the cooled nozzle in the current hydrogen 
atomic beam experiment~\cite{NiEtAl2000,FiEtAl2004}, 
the slow hydrogen atoms enter a high vacuum 
which is kept at room temperature.

\begin{table*}
\begin{center}
\begin{minipage}{0.9\linewidth}
\begin{center}
\caption{\label{table2} Black-body energy shift of the 
$2S$ excited state of atomic hydrogen and ionized atomic helium,
evaluated according to Eq.~(\ref{PnS}), 
expressed in terms of frequencies via division by the 
Planck constant $h$. Note that the 
entries for $\Delta E'_{\mathrm{bb}}( 2S ) + 
\Delta E''_{\mathrm{bb}}( 2S )$ are given here only as an indication
of the quality of the approximation given in Eqs.~\eqref{Deltae2Sbbprime} 
and~\eqref{Deltae2Sbbdprime} and therefore in brackets. The entries 
for $\Delta E_{\mathrm{bb}}( 2S )$ are obtained by numerical
integration of \eqref{ebb} and thus relevant for comparison 
to experiment. The explanation for the discrepancy of the approximation
and the numerical integration for 
$4\,{\mathrm K}$ and $Z=2$ is given in the text
(see also Fig.~\ref{fig1}).}
\begin{tabular}{c@{\hspace{0.5cm}}cr@{\hspace{0.0cm}}l@{\hspace{0.6cm}}%
r@{\hspace{0.0cm}}l}
\hline
\hline
\multicolumn{1}{c}{\rule[-2mm]{0mm}{6mm} Nucl. charge number} &
\multicolumn{1}{c}{Temperature} &
\multicolumn{2}{c}{$h^{-1} \Delta E_{\mathrm{bb}}( 2S )$} & 
\multicolumn{2}{c}{$h^{-1} \, [\Delta E'_{\mathrm{bb}}( 2S ) + 
\Delta E''_{\mathrm{bb}}( 2S )]$} \\
\hline
\rule[-2mm]{0mm}{6mm}
$Z = 1$ &
$4 \, {\rm K}$ &
$7.$&$79 \times 10^{-7}\, {\mathrm{Hz}}$ &
($8.$&$13 \times 10^{-7}\, {\mathrm{Hz}}$) \\
\rule[-2mm]{0mm}{6mm}
$Z = 1$ & 
$77 \, {\rm K}$ & 
$-1.$&$44 \times 10^{-3} \, {\mathrm{Hz}}$ &
($-1.$&$46 \times 10^{-3} \, {\mathrm{Hz}}$) \\
\rule[-2mm]{0mm}{6mm}
$Z = 1$ & 
$300 \, {\rm K}$ & 
$-9.$&$89 \times 10^{-1}\, {\mathrm{Hz}}$ &
($-9.$&$87 \times 10^{-1} \, {\mathrm{Hz}}$) \\
\rule[-2mm]{0mm}{6mm}
$Z = 2$ &
$4 \, {\rm K}$ & 
$3.$&$40 \times 10^{-6}\, {\mathrm{Hz}}$ &
($3.$&$30 \times 10^{-5}\, {\mathrm{Hz}}$) \\
\rule[-2mm]{0mm}{6mm}
$Z = 2$ &
$77 \, {\rm K}$ & 
$1.$&$18 \times 10^{-2}\, {\mathrm{Hz}}$ &
($1.$&$19 \times 10^{-2}\, {\mathrm{Hz}}$) \\
\rule[-2mm]{0mm}{6mm}
$Z = 2$ &
$300 \, {\rm K}$ & 
$1.$&$21 \times 10^{-1}\, {\mathrm{Hz}}$ &
($1.$&$21 \times 10^{-1}\, {\mathrm{Hz}}$) \\
\hline
\hline
\end{tabular}
\end{center}
\end{minipage}
\end{center}
\end{table*}

%
%
\section{Imaginary Part of Black--Body Induced Energy Shift}
\label{imagparts}

In order to evaluate the black-body induced 
decay rates, we have to be more precise than in 
Eqs.~(\ref{pol}) and (\ref{ebb}). 
Namely, we have to collect all the poles from the 
bound-state spectrum as well as integrate over the 
transitions to the continuous spectrum,
and introduce infinitesimal imaginary displacement into 
to propagator denominators. The relevant formulas are
\begin{subequations}
\begin{align}
\label{poleps}
& P_{\nu}(nS) \nonumber\\
& = 
\frac13 \, 
\sum_{i=1}^3 
\left( \left< nS \left| x^i \frac{1}{H - \ii \epsilon - E(nS) - h \nu} x^i
\right| nS \right> \right.
\nonumber\\
& \left. \qquad +
\left< nS \left| x^i \frac{1}{H - \ii \epsilon - E(nS) + h \nu} x^i
\right| nS \right> \right)
\,,
\\[2ex]
\label{ebbeps}
& \Delta {\mathcal E}_{\mathrm{bb}}( nS ) = - 
\lim_{\epsilon \to 0^+} \frac{ e^2 }{2} \,
\int\limits_0^\infty \dd \nu\,
{\vec e}^{\,2}(\nu) \, P_{\nu}(nS) 
\nonumber\\
& \qquad \quad =
\Delta E_{\mathrm{bb}}( nS ) -  \ii \, 
\frac{ \Delta \Gamma_{\mathrm{bb}}( nS ) }{2} \,.
\end{align}
\end{subequations}
The first term on the right-hand side of (\ref{poleps})
describes a process with an absorption of a photon 
by the atom from a black-body mode,
which lowers the energy of the black-body photon field
in the virtual state (term $-h \nu$),
and a subsequent emission into a black-body mode.
For this term, the imaginary part is generated for virtual atomic states,
as contained in the spectral decomposition of $H$,
which have an energy higher than that of the reference state $E(nS)$.
This is the only relevant process to generate a black-body induced
width if the ground state acts as a reference state,
and it is easy to see that the imaginary part of the 
energy shift thus generated has the correct sign, i.e.~it 
contributes a negative imaginary part to 
$\Delta {\mathcal E}_{\mathrm{bb}}( nS )$.

The second term on the right-hand side of (\ref{poleps})
describes a process with (first) emission into a black-body mode,
then absorption from the black-body field.
In this case, the imaginary part is generated for virtual atomic states
of an energy lower than that of the reference state $E(nS)$.
This process can be relevant for the generation of a black-body induced
decay when the $2S$ states acts as a reference state,
and the virtual state is a $2P_{1/2}$ state.
It is again easy to see that the imaginary part of the
energy shift thus generated has the correct sign, i.e.~it also
contributes a negative imaginary part to
$\Delta {\mathcal E}_{\mathrm{bb}}( nS )$.

We can then easily obtain the black-body induced decay 
rates of the $1S$ and $2S$ states.
Numerical results are given in Table~\ref{table3},
where we restrict ourselves to atomic hydrogen.
There is a very strong increase of the rates with temperature,
and we note a comparatively large induced decay rate for the 
$2S$ state, due to the proximity of the $2P_{1/2}$ and 
$2P_{3/2}$ levels.

\begin{table}
\begin{center}
\begin{minipage}{0.9\linewidth}
\begin{center}
\caption{\label{table3} Black-body induced width 
$\Delta \Gamma_{\rm bb}(1S)/h$ of the ground state and of the 
$2S$ excited state,
evaluated according to Eq.~(\ref{ebbeps}) for atomic hydrogen. The 
results are given in Hz (cycles per second).}
\begin{tabular}{ccr@{\hspace{0.0cm}}l}
\hline
\hline
\multicolumn{1}{c}{\rule[-2mm]{0mm}{6mm} State} &
\multicolumn{1}{c}{Temperature} &
\multicolumn{2}{c}{$h^{-1} \Delta \Gamma_{\mathrm{bb}}( nS )$} \\
\hline
\rule[-2mm]{0mm}{6mm}
$1S$ &
$300 \, {\rm K}$ &
$1.$&$13 \times 10^{-163}\; {\mathrm{Hz}}$ \\
\rule[-2mm]{0mm}{6mm}
$1S$ &
$3000\, {\rm K}$ & 
$2.$&$15 \times 10^{-9}\; {\mathrm{Hz}}$ \\
\rule[-2mm]{0mm}{6mm}
$1S$ &
$30000 \, {\rm K}$ & 
$8.$&$00 \times 10^6\; {\mathrm{Hz}}$ \\
\rule[-2mm]{0mm}{6mm}
$2S$ &
$300 \, {\rm K}$ & 
$3.$&$08 \times 10^{-3}\; {\mathrm{Hz}}$ \\
\rule[-2mm]{0mm}{6mm}
$2S$ &
$3000 \, {\rm K}$ & 
$7.$&$49  \times 10^{3}\; {\mathrm{Hz}}$ \\
\rule[-2mm]{0mm}{6mm}
$2S$ &
$30000 \, {\rm K}$ & 
$2.$&$34 \times 10^{7}\; {\mathrm{Hz}}$ \\
\hline
\hline
\end{tabular}
\end{center}
\end{minipage}
\end{center}
\end{table}

%
%
\section{Conclusions}
\label{conclu}

In this paper, we have set up
a formalism for the treatment of black-body induced energy shifts
and corresponding decay rates (widths) for low-lying atomic 
states in atomic systems with a low nuclear charge.
We have evaluated our expressions for the $1S$ and $2S$ states
of atomic hydrogen and hydrogenlike (ionized) helium.
The formula~(\ref{PnS}) leads to a consistent treatment 
of the virtual $nP_j$ states ($j=3/2$ and $j=1/2$)
which are displaced from the 
reference $nS_{1/2}$ state only by a fine-structure splitting 
and by the Lamb shift, respectively (see Sec.~\ref{realparts}),
and can be easily generalized to other cases of interest.
We note that all black-body shifts given in Tables~\ref{table1}
and~\ref{table2} are well below 1~Hz in absolute magnitude
and do not have to be taken into account at current and 
projected levels of accuracy for high-precision laser spectroscopy,
although it is perhaps useful to remark that they are larger
than other non-standard effects (``accuracy limits'') for two-photon
spectroscopy recently discussed in the literature 
(Refs.~\cite{JeMo2002,LaScSoPl2007}).

The imaginary part of the black-body induced energy shift
(``black-body ac Stark shift'') has been discussed and evaluated in 
Sec.~\ref{imagparts}. The corresponding decay rates 
given in Table~\ref{table3} can be quite substantial, 
especially at elevated temperatures. At $T = 3000 \, {\rm K}$, 
the $2S$ state of hydrogen acquires a black-body width 
of about 8~kHz, and at $T = 30000 \, {\rm K}$, the ground-state
of atomic hydrogen is roughly 8~MHz wide, purely due to the black-body 
induced interactions. At the same temperature, the 
Boltzmann factor $\exp[-(E_{2P}-E_{1S})/(k_B T)]$ for the 
excitation of the ground state into the $2P$ state is only 2\%.

Although the black-body energy shifts have been studied 
quite intensively, the numerical results and approximations
have not yet appeared in the literature to the best of our knowledge.
We have worked in SI units throughout this article in order
to enhance the clarity of the details of the derivation.

%
%
\section{Acknowledgments}

U.D.J.~acknowledges support by the Deutsche Forschungsgemeinschaft
(contract Je285/3--1).

\end{document}